\documentclass[preprint,superscriptaddress,longbibliography,onecolumn,floatfix,showpacs,amsmath,amssymb,amsfonts]{revtex4-2} 

\usepackage{graphicx}

\newcommand{\mi}{\textrm{i}} 
\newcommand{\me}{\mathrm{e}}

\usepackage{siunitx} 
\DeclareSIUnit{\rad}{rad}
\usepackage[normalem]{ulem}
\usepackage{textcmds}

\newcommand{\freiburg}{Institute for Theoretical Physics, ETH Zürich, CH-8093 Zürich, Switzerland}
\newcommand{\kassel}{Institut f\"{u}r Physik, Universit\"{a}t Kassel, Heinrich-Plett-Stra{\ss}e 40, 34132 Kassel, Germany}
\newcommand{\ethZ}{Institute of Quantum Electronics, ETH Z\"{u}rich, Auguste-Piccard-Hof 1, 8093 Z\"{u}rich, Switzerland}
\newcommand{\madrid}{Departamento de F\'{i}sica Te\'{o}rica de la Materia Condensada and Condensed Matter Physics Center (IFIMAC), Universidad Aut\'{o}noma de Madrid, E-28049 Madrid, Spain}

\begin{document}

\title{Experimentally separating\\ vacuum fluctuations from source radiation}

\author{Alexa Herter}
\altaffiliation{Corresponding authors}
\affiliation{\ethZ}
\author{Frieder Lindel}
\affiliation{\freiburg}
\affiliation{\madrid}
\author{Laura Gabriel}
\affiliation{\ethZ}
\author{Stefan Yoshi Buhmann}
\affiliation{\kassel}
\author{J\'{e}r\^{o}me Faist}
\altaffiliation{Corresponding authors}
\affiliation{\ethZ}

\date{\today}

\maketitle


\textbf{The unique distinction between vacuum-field and source-radiation induced effects in processes such as the Lamb shift, Casimir forces or spontaneous emission, remains unresolved even at the theoretical level, and an experimental approach was never considered feasible \cite{Milonni2013TheVacuum,Milonni1973InterpretationEmission, Senitzky1973Radiation-ReactionElectrodynamics, Dalibard1982VacuumContributions}. In 1932, Fermi introduced the two-atom problem, which is a Gedanken-experiment that explores how two atoms interact with the surrounding electromagnetic field via vacuum and source-radiation induced processes, providing fundamental insights into the behavior of quantum fields \cite{Fermi1932QuantumRadiation,Biswas1990VirtualAtoms, Valentini1991Non-localElectrodynamics, Milonni1995PhotodetectionOptics, Pozas-Kerstjens2015HarvestingVacuum}. Recent advancements in ultrafast optics have enabled experimental analogues of this system using two laser pulses inside a nonlinear crystal \cite{Riek2015DirectFluctuations,Benea2019,Settembrini2022DetectionCone}.
Here, we demonstrate the detection of vacuum and source radiation induced correlations, separated by their causal properties, between two laser pulses. 
In particular, we show that vacuum fluctuations and source radiation correlate different quadratures of near-infrared laser pulses, allowing them to be individually probed through phase-sensitive detection. 
This result provides an experimental verification of the time-domain fluctuation-dissipation theorem at the quantum level and offers a novel path to studying quantum radiation effects in time-dependent media. Beyond resolving a longstanding theoretical ambiguity, our findings open new possibilities for investigating quantum field phenomena in the context of relativistic quantum information such as entanglement harvesting from the quantum vacuum or quantum field detection in analogues of curved space-times.}\\

\begin{figure}
    \centering
    \includegraphics[width=9cm]{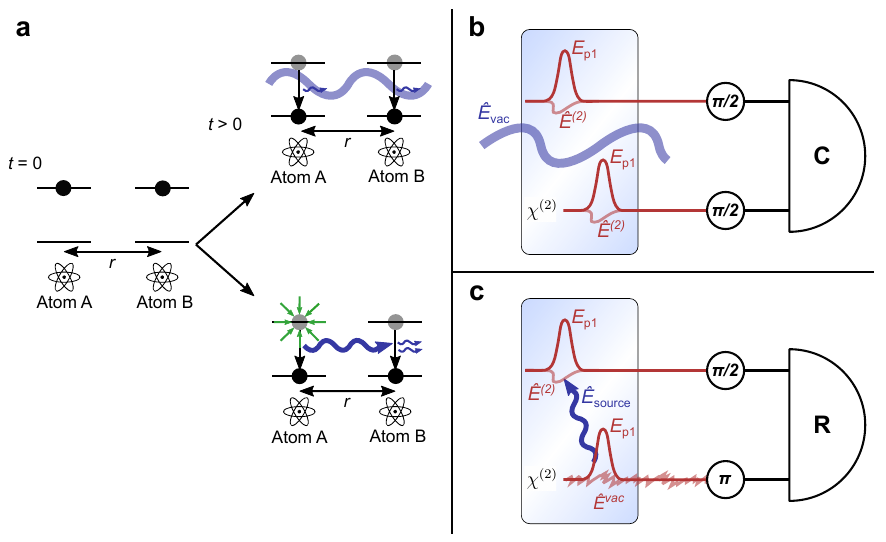}
    \caption{\textit{a: Fermi's two-atom problem.} 
    Two initially uncorrelated excited atoms A and B are separated by a distance $r$ and their interaction with the electro-magnetic field is turned on at time $t=0$.  The atoms can become correlated either by interacting individually with the surrounding vacuum field fluctuations (upper sketch), or by exchanging source radiation (one atom emitting a photon by interacting with its own radiation field, which then interacts with the other atom) (lower sketch).
    \textit{b: Detecting vacuum induced correlations.} Similar to the vacuum-induced correlations in Fermi's two-atom setup, the polarization state of two laser pulses can become correlated by interacting with the surrounding vacuum field inside an electro-optic crystal. The generated nonlinear field is detected in a homodyne detection scheme \cite{Moskalenko2015ParaxialVacuum,Benea2019}.
    \textit{c: Detecting  correlations induced by source radiation.} Additionally, one laser pulse can emit a photon due to the interaction with the ground-state fluctuations in its own frequency range and the generated photon can then interact with the second pulse. To probe this process, the phase relation in one of the homodyne detection schemes has to be changed from $\frac{\pi}{2}$ to $\pi$ \cite{Lindel2023ProbingTime} (see main text).}
    \label{fig:two-atom}
\end{figure}

Fermi's two atom setup consists of two atoms in empty space whose interaction with the vacuum radiation field is abruptly switched on. Studying the subsequent build up of correlations between the atoms reveals fundamental properties of quantum fields in space and time \cite{Biswas1990VirtualAtoms, Valentini1991Non-localElectrodynamics, Milonni1995PhotodetectionOptics,Pozas-Kerstjens2015HarvestingVacuum}.
In the Heisenberg picture, two different terms contribute to these correlations \cite{Milonni2013TheVacuum,Dalibard1982VacuumContributions,Lindel2023ProbingTime,Tjoa2021WhenHarvesting}, as illustrated in Fig.\,\ref{fig:two-atom}\,\textbf{a}. On one hand, correlations between the two atoms arise from their individual interaction with the ground state electro-magnetic field -- the \textit{vacuum fluctuations} -- whereby correlations intrinsically existing in the vacuum field are swapped to the atoms \cite{Biswas1990VirtualAtoms, Valentini1991Non-localElectrodynamics, Milonni1995PhotodetectionOptics, Pozas-Kerstjens2015HarvestingVacuum}. 
On the other hand,  the two atoms can exchange a real photon. This process, triggered by the interaction of the emitting atom with its own radiative field (radiation reaction) is known as \textit{source radiation} \cite{ Biswas1990VirtualAtoms, Dalibard1982VacuumContributions}. 
Mathematically, the individual contributions of vacuum fluctuations and source radiation are in general not uniquely determined \cite{Milonni1973InterpretationEmission, Senitzky1973Radiation-ReactionElectrodynamics, Dalibard1982VacuumContributions}, since they depend on the initial ordering of two commuting operators.

While for most of the phenomena associated with interactions with the ground state of light, vacuum fluctuations and source radiation can therefore be seen as inseparable  "two sides of the same quantum mechanical coin" \cite{Senitzky1973Radiation-ReactionElectrodynamics}, 
they have a different character regarding causality in Fermi's two-atom problem: The vacuum-induced correlations arise instantaneously upon turning on the interaction, while the exchange of a photon leads to correlations only after the propagation time $\delta t=\frac{r}{c}$ \cite{Lindel2023ProbingTime,Tjoa2021WhenHarvesting}.  
However, different choices of operator ordering can lead to a non-causal description of source radiation, making its interpretation ambiguous. 
This issue remains purely theoretical unless the two contributions can be observed independently.  
Yet, an experimental investigation has long been considered unfeasible due to the challenge of initiating the two atoms, rapidly switching their interaction with the field on ultrashort timescales, and detecting the resulting correlations \cite{Sabin2011FermiQED}. Here, we overcome these limitations using an all optical version of Fermis two-atom setup and probe source radiation and vacuum field fluctuations individually for the first time. \\


In contrast to atoms, the interaction of laser pulses with a surrounding electro-magnetic field can be initiated (terminated) on ultra-short time-scales by the laser pulses entering into (propagating out of) a nonlinear material. This has been exploited in a recent experiment to demonstrate the mapping of the vacuum-induced correlations between two atoms in Fermis two-atom setup onto two spatially separated laser pulses coupled to the vacuum field inside a nonlinear crystal \cite{Settembrini2022DetectionCone}, compare Fig.~\ref{fig:two-atom}b. Specifically, the technique of electro-optic sampling was used, where the interaction between near-infrared laser pulses and a terahertz or mid-infrared field (possibly in the vacuum state) generates a second order nonlinear field polarized perpendicular to the original laser pulse \cite{Moskalenko2015ParaxialVacuum,Riek2015DirectFluctuations,Benea2019}. This leads to a change in the state of the two near-infrared pulses emerging from the crystal, corresponding to the transition in the atoms in Fermi's two-atom setup.
Recently, a similar experiment has been theoretically suggested \cite{Lindel2023ProbingTime}, which would be capable of individually detecting correlations caused by source radiation as well (see Fig.~\ref{fig:two-atom}c), and in the following we demonstrate its experimental implementation.


The idea relies on the fact that in the electro-optic sampling analogue of Fermi's two-atom setup, vacuum-field fluctuations (given by the correlation function of the THz vacuum field $\mathbf{\mathsf{C}}(\mathbf{r},\mathbf{r}',t,t')
=\tfrac{1}{2}\langle\{\hat{\mathbf{E}}_\mathrm{vac}(\mathbf{r},t),\hat{\mathbf{E}}_\mathrm{vac}(\mathbf{r}',t)'\}\rangle$) and source radiation (given by the response function $\mathbf{\mathsf{R}}(\mathbf{r},\mathbf{r}',t,t') =(\mathrm{i}/\hbar)\theta(t-t')\langle [\hat{\mathbf{E}}_\mathrm{vac}(\mathbf{r},t),\hat{\mathbf{E}}_\mathrm{vac}(\mathbf{r}',t')]\rangle$) correlate different quadratures of the two near-infrared pulses, which allows to individually probe them by changing the phase plate in the employed homodyne detection scheme (see Fig.~\ref{fig:two-atom}b,c). In the vacuum-field contribution, the near-infrared field of the two pulses individually interact with the vacuum field at terahertz frequencies. Due to the retardance of $\frac{\pi}{2}$ occurring in the nonlinear process, the field quadrature orthogonal to the quadrature of the laser pulses are correlated. These correlations can be accessed via quarter-wave plates in the homodyne detection scheme. 
The instantaneous correlation of the two pulses induced by the terahertz vacuum fluctuations are obtained from the individually detected signals as a function of different time-delays between the two probe pulses \cite{Settembrini2022DetectionCone}. 
In the source radiation contribution, the electric field of one probe pulse interacts with the quadrature of the co-propagating ground-state fluctuations in the near infrared that is in-phase with the pulse, resulting in the emission of source radiation. 
The second pulse is then interacting with that terahertz source radiation in the same way as with the terahertz vacuum field again leading to a nonlinear field contribution that has a quadrature shifted by $\pi/2$ compared to the pulse. 
To observe the correlations resulting from source radiation, one thus has to probe the first pulse with a half-wave plate---thereby accessing the quadrature that is in-phase with the pulse---and the second with a quarter-wave plate as in the vacuum field contribution (see Ref.~\cite{Lindel2023ProbingTime} for details). \\


\begin{figure}
    \centering
    \includegraphics[width=9cm]{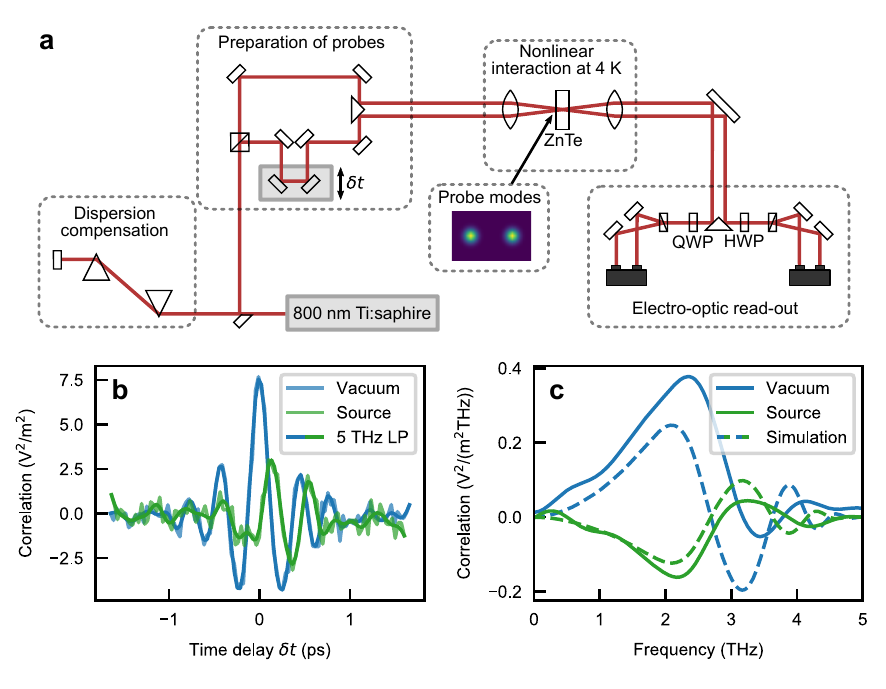}
    \caption{\textbf{a}: \textit{Experimental setup to measure electro-optic correlations.} A prism compressor is used to compensate for the dispersion of the transmission optics in the setup to achieve \SI{110}{\fs} long pulses inside the nonlinear crystal. Two individual probe signals are prepared whose time-delay is controlled using a delay stage. Inside a cryostat, the probe signals are focused tightly into the zinc-telluride crystal (ZnTe) at \SI{4}{\K}. Afterwards the polarization change on both probe beams is individually evaluated on a balanced detector. Depending on the experiment, a quarter-wave plate (QWP) is used to access the change in ellipticity or a half-wave plate (HWP) to observe a rotation in the polarization. \textbf{b}: \textit{Temporal electro-optic field correlation.} The correlations caused by terahertz vacuum fluctuations (blue) and source radiation (green) is plotted depending on the time-delay between the two pulses $\delta t$. The opaque green and blue curves show the original data, while the faint lines show the data after applying a \SI{5}{THz} low-pass filter to suppress electronic noise. \textbf{c}: \textit{Electro-optic correlations in frequency domain.} Experimentally observed correlation are shown obtained from the Fourier-transformation of the time-traces shown in \textbf{b}. The vacuum-induced correlation is plotted in blue and green is the imaginary part of the correlation caused by source radiation. The simulated spectral correlations (see Methods and Ref.~\cite{Lindel2023ProbingTime}) are plotted with dashed lines.}
    \label{fig:results}
\end{figure}
As shown in Fig.\,\ref{fig:results}\,\textbf{a},  the probe signal is pre-stretched in a prism compressor before entering the electro-optic detection setup similar to Ref.~\cite{Settembrini2022DetectionCone} to compensate for dispersion and obtain \SI{110}{\fs}-short pulses inside the detection crystal (see Methods and Extended Data Fig. 1). 
In agreement with Ref.~\cite{Settembrini2022DetectionCone}, the vacuum-induced correlation for the two beams separated by $\SI{50}{\um}$ and measured with a quarter-wave plate on both detection lines oscillates symmetrically around zero time delay $\delta t$ with the maximum correlation for $\delta t = 0$ (blue line in Fig.\,\ref{fig:results}\,\textbf{b}). 
Compared to previous results, the reduced pulse duration increases the observed peak correlation to more than \SI{5}{\V^2\per\m^2} \cite{Benea2019,Settembrini2022DetectionCone}. 


For the setup sensitive to correlations stemming from source radiation (combination of a half-wave plate on one and a quarter-wave plate on the other detection line), the temporal correlations (green line in Fig.\,\ref{fig:two-atom}\,\textbf{b}) vanish for zero time delay $\delta t=0$ . The signal occurs mainly for positive time delays $\delta t > 0$, which means the pulse generating the source radiation arrives at the crystal before the pulse detecting it, confirming the causal nature of correlations induced by source radiation. 


The Fourier transform of the correlation traces for vacuum fluctuations and source radiation are plotted in Fig.\,\ref{fig:results}\,\textbf{c}. Compared to our previous results, the signal bandwidth is now extending to \SI{4.5}{THz} thanks to the shorter pulse length \cite{Settembrini2022DetectionCone}. 
The initial increase of the signal with frequency up to \SI{2}{THz} results from the increasing strength of the vacuum fluctuations with frequency.
Above \SI{2.5}{THz} the two signals change sign as the distance between the two beams becomes comparable to the wavelength inside the crystal.


The vacuum-field and source radiation signals are further not independent from each other but connected via the fluctuation-dissipation theorem. It predicts a phase-shift of $\frac{\pi}{2}$ between the temporal correlation caused by source radiation and vacuum fluctuations \cite{Pottier2001QuantumFormulation,Lindel2023ProbingTime}, which is clearly visible in the experimental data in Fig.~2b. This makes the experimental setup a probe of the \emph{time-domain} fluctuation-dissipation theorem in the quantum limit (zero temperature).
Moreover, the imaginary part of the spectral correlation caused by source radiation (green) is compared to the vacuum-induced spectral correlation (blue) obtained in the experiment (solid lines) and in simulations (dashed lines) in Fig.\,\ref{fig:results}\,c. 
Here, the fluctuation-dissipation theorem dictates that the imaginary part of the source radiation signal is equivalent to the vacuum-induced correlations multiplied by a factor $-\frac{1}{2}$ \cite{Lindel2023ProbingTime}.
The slight deviation between the two curves is caused by the uncertainty of the beam position inside the focal plane. \\


We have presented an individual experimental detection of correlations arising between two laser pulses due to vacuum fluctuations and source radiation naturally separated by their causal properties. Dalibard et al. \cite{Dalibard1982VacuumContributions} proposed already in 1982 a way to lift the ambiguity between source radiation and vacuum fluctuations that is consistent with ours by demanding the operators describing the two contributions to be each Hermitian and therefore physical observables. Our results hence confirm the physical meaning of such a separation and support Dalibard’s argument. While causality enables the separation of vacuum fluctuations and source radiation in Fermi’s two-atom problem, Dalibard’s method, now validated by our experiment, provides a general approach applicable to any effect involving these fundamental quantum contributions.

Furthermore, the measurements agree with the predictions of the time-domain fluctuation-dissipation theorem as expected from the discussion of the two-atom problem \cite{Lindel2023ProbingTime}. 
The direct connection of the spectral correlations induced by source radiation and vacuum fluctuations offers a new access to vacuum fluctuations through the measurement of source radiation and vice versa.


So far the experiment has been performed at \SI{4}{\K} to suppress any thermal radiation in the terahertz regime. 
While prior studies have demonstrated the electro-optic correlation induced by thermal photons at room temperature \cite{Benea2019,Settembrini2024ThirdFrequencies}, the correlation caused by source radiation is anticipated to be independent of the optical state in the THz regime \cite{Lindel2023ProbingTime,Tjoa2021WhenHarvesting}.
Consequently, the configuration that has been demonstrated, which utilizes a half-wave plate and a quarter-wave plate, possess the capability to detect the correlations caused by a single THz photon within a thermal photon pool at room temperature. 


Furthermore, the experiment demonstrates time-domain access to quantum fields at the single photon level. This offers new experimental possibilities for the study of quantum radiation phenomena in time-varying media \cite{Galiffi2022PhotonicsMedia}, such as the dynamical Casimir effect \cite{Moore1970QuantumCavity}, or in relativistic quantum information theory \cite{Fuentes-Schuller2005AliceFrames,Mann2012RelativisticInformation}, e.g. for the experimental demonstration of entanglement harvesting from the vacuum state \cite{Lindel2024EntanglementFields} or for probing quantum correlations of relativistic fields in analogue curved space-times  \cite{Kizmann2019SubcyclePerspective}, including event horizons \cite{Philbin2008Fiber-opticalHorizon}.\\

\newpage

\bibliographystyle{paper}
\bibliography{references}

\listoffigures
\section*{Methods} 
\subsection*{Experimental setup}
The optical setup to detect electro-optic correlations is shown schematically in Fig.\,\ref{fig:results}\,\textbf{a} of the main text.
We are using the pulsed signal of a titan-sapphire laser centered at \SI{800}{\nm} to generate the probe signal. To compensate for the dispersion caused by several transmission optics, we implemented a prism compressor. The beam position is actively stabilized to reduce long-term drifts. A beam splitter divides the intensity of the pulsed signal equally into two beam paths and a delay stage controls the time delay $\delta t$ between the two beams. 
The two beams are placed in close proximity to each other and two piezo-controlled mirrors set the angle between the two beams. Afterwards the two beams are directed inside a dilution fridge. 

Inside the cryostat, a \SI{50}{\mm} lens focuses the beams into a $\langle110\rangle$-cut zinc-telluride crystal. The angle of the two beams determine the distance of the two beams in the focal plane and are adjusted to a distance of \SI{50}{\um} in the current experiment. Since the beam waist measures \SI{10}{\um}, the two beams are well-separated inside the nonlinear crystal. A second lens re-collimates the beams afterwards. Both lenses and the nonlinear crystal are thermally connected to the \SI{4}{\K}-plate of the cryostat to suppress thermal radiation in the terahertz-regime. The two beams are propagating along the $\langle\bar1\bar10\rangle$-axis of the zinc telluride and are polarized along the $\langle 001\rangle$-direction.

After leaving the cryostat, the two probe beams are separated and the change in polarization is analyzed individually using the combination of a wave plate, a polarizing beam splitter and a balanced detector. The correlation of the two signals is calculated for each individual pulse pair using a fast acquisition card. Further details about the setup and data acquisition, which is suppressing slowly-varying correlations, are reported in \cite{Benea2019}. 

\subsection*{Pulse duration inside the detection crystal}
To control and optimize the pulse duration of the probe pulses inside the zinc-telluride crystal, a prism-compressor using two SF10-prisms in the Brewster configuration has been added to the optical beam path. 
Depending on the lateral position of the prisms with respect to the beam path, this compressor can add a group delay dispersion of about \SI{-1500}{\fs^2} up to \SI{-6000}{\fs^2}.
A third-order nonlinear mixing process between the two beams overlapping inside the crystal is used to measure the temporal intensity correlation function of the probe signal and minimize the pulse duration inside the zinc-telluride crystal adjusting the position of the prisms of the compressor \cite{Caumes2002Kerr-LikeCrystals,Tian2008QuantitativeZnTe,Settembrini2024ThirdFrequencies}. 
In the Extended Data Fig.\,1 the detected temporal intensity correlation with a full-width half-maximum of \SI{168}{\fs} is shown. Assuming a sech$^2$-shaped pulse, we determine an intensity full-width half-maximum of the probe signal of \SI{110}{\fs} compared to \SI{200}{\fs} in previous experiments not using dispersion compensation \cite{Settembrini2022DetectionCone,Settembrini2024ThirdFrequencies}.

\subsection*{Determining the zero-time delay position}
To judge the causality of source radiation, but also to separate the spectra into real and imaginary part, it is crucial to know the position of the zero-time delay. 
At the same time, due to the long beam path until the pulses reach the crystal inside the cryostat (further information in reference \cite{Benea2019}), slight changes in the alignment can shift its position. 

We apply RF-referencing (described in Ref. \cite{Benea2019}) to suppress any correlation occurring from slow drifts in the optical setup or coherent nonlinear mixing processes between the two probe signal. 
To estimate the position of the zero-time delay we use the correlation without applying this filter, which is acquired as well. 
During the detection of vacuum fluctuations, when two quarter-wave plates are analyzing the polarization change, a peak occurs in the raw data (i.e. before RF referencing), when the two beams overlap in time, probably caused by a coherent nonlinear mixing process between the two non-vanishing tails of the Gaussian spatial modes. 
As shown in the Extended Data Fig.\,2\,\textbf{a} the peak position changes between two neighboring stage positions during the acquisition of 94 traces measuring the vacuum-induced correlation. \\

From the Fourier transformation of the vacuum-induced correlation assuming the temporal overlap position at \SI{10.775}{\mm} we can determine the remaining temporal shift. 
Due to the symmetry of the experimental configuration, we expect a constant phase of the spectral correlation. Fitting the phase as shown in the Extended Data Fig.\,2\,\textbf{b} between \SI{1.7}{\THz} and \SI{2.7}{\THz}, where the correlation is strong enough to extract the slope of the phase, reveals a temporal shift of \SI{-0.01}{\ps} corresponding to changes in stage position of \SI{-1.5}{\um}. In the Extended Data Fig.\,2\,\textbf{c} the real and imaginary part of the vacuum-induce spectral correlation is shown with the zero-time delay determined this way. The blue and orange shaded area indicates the uncertainty assuming a time-shift of $\pm \SI{16.7}{\fs}$ corresponding to the temporal distance between two neighboring data points.\\

Due to the usage of a half-wave plate on one of the two detection lines, neither the symmetry of the signal, nor the non-RF reference signal can be used to determine the zero-time delay position. 
Therefore we measure the correlation due to source radiation and vacuum fluctuation right after each other without realigning the setup to be able to assume the same timing in the two measurements. 
The exchange of the wave plate does not influence the timing of the pulses, since they are placed in the beam path after the nonlinear crystal. \\

Evaluating the signals on the individual detectors, we observe a strong drift after acquiring 100 traces, which led to the saturation of one detector. Within trace 105 the detector returned to a non-saturated operation. 
As shown in the Extended Data Fig.\,2\,\textbf{d}, averaging separately the traces from 0 to 100 and from 105 to the end and applying a 5-THz low pass to suppress high-frequency noise reveals a drift in the zero-time delay in the time in between. To determine the shift precisely, the averaged traces shown in the Extended Data Fig.\,2\,\textbf{d} are interpolated using a cubic spline to increase the temporal resolution. The cross-correlation between these two curves reveals a time-shift of \SI{50.2}{\fs} as shown in the Extended Data Fig.\,2\,\textbf{e}. After shifting one of the two traces by one temporal step of \SI{33.3}{\fs}, the weighted average between the unfiltered trace before and after the incident is calculated to determine the final time trace.\\

For the final Fourier transformation shown in Fig.\,2\,\textbf{f}, the traces are transformed into the frequency domain individually and averaged after applying the phase-factor correcting for the respective time delay. The shaded area indicates the uncertainty due to the complex phase assuming a temporal uncertainty interval of $\pm\SI{16.7}{\fs}$.

\subsection*{Beam distance in vacuum-induced correlation}
As shown in Fig.\,2\,\textbf{b} of the main text, the spectral correlation induced by vacuum fluctuations deviates from the simulated signal. Besides an higher signal strength than expected, the change between positive and negative correlation is shifted towards higher frequencies. 
Since the switch between correlation and anti-correlation depends on the distance between the two focal spots of the laser pulses, we simulated the vacuum-induced correlation additionally for smaller distances. 
In Fig.\,3\,\textbf{a}, the spectral correlation simulated for a beam distance of \SI{30}{\um} reproduces well the signal observed in the measurement. For further confirmation, the measured temporal correlation is compared to the simulation assuming a beam distance of \SI{30}{\um} in Fig.\,3\,\textbf{b}. To move the two focal spots from an overlapping position to a distance of \SI{50}{\um}, the angle of the beams entering the dilution fridge is changed by \SI{0.5}{\milli\rad}. During a total acquisition time of several weeks, a change in the beam angle of a few tens of mrad is within the expected experimental uncertainty. 

\subsection*{Theoretical model}

To model the experiments, we use the theoretical framework of two-beam electro-optical sampling developed in Ref.~\cite{Lindel2023ProbingTime}. It builds on previous works in Refs.~\cite{Moskalenko2015ParaxialVacuum,Lindel2020TheoryDetection}.

Applying the paraxial wave-approximation to all involved near-infrared fields but not the THz field \cite{Lindel2020TheoryDetection}, the electro-optic sampling signal with two quarter wave plates reads \cite{Lindel2023ProbingTime}
\begin{align} \label{eq:EOSSignalVF}
G_{\mathrm{vac}}(\delta t)  = \int\limits_{V_C} \mathrm{d}\mathbf{r} \int \mathrm{d} t \int \limits_{V_C} \mathrm{d}\mathbf{r}^\prime \int   \mathrm{d} t^\prime  L_1(\mathbf{r}, t)L_2(\mathbf{r}^\prime, t^\prime)C(\mathbf{r}, \mathbf{r}^\prime, t,t^\prime) .
\end{align} 
With a quarter wave plate for pulse one and a half-wave plate for pulse two it is in turn given by
\begin{align}\label{eq:EOSSignalSR}
G_\mathrm{s}(\delta t)  = -  \frac{\hbar}{2} \int\limits_{V_C} \mathrm{d}\mathbf{r} \int \mathrm{d} t \int \limits_{V_C} \mathrm{d}\mathbf{r}^\prime \int   \mathrm{d} t^\prime L_{1}(\mathbf{r}, t)L_{2}(\mathbf{r}^\prime, t^\prime)   R(\mathbf{r}, \mathbf{r}^\prime, t,t^\prime).
\end{align} 
Here, $V_C$ is the volume of the nonlinear crystal and $L_1(\mathbf{r},t) = L_{2}(\mathbf{r}-\delta r\mathbf{e}_x, t-\delta t) = (2/\pi)^{3/2}\me^{-2(t-n_gz/c)^2/\tau_\sigma^2} \me^{-2r_\parallel^2/w^2}$ [$\mathbf{r}_\parallel = (r_x, r_z)$, $w$: beams waist, $\tau_\sigma$: pulse duration, $n_g$: group refractive index in the near-infrared] are the Gaussian shaped pulse envelopes of the two laser pulses. The response and correlation functions of the $x$-polarized THz field $\hat{E}(\mathbf{r},t)$ inside the crystal accounting for dispersion and absorption are defined in the main text and can be evaluated using macroscopic quantum electrodynamics \cite{Scheel2008MacroscopicApplications}. In frequency domain, they are given by $ R(\mathbf{r},\mathbf{r}^\prime, \Omega) = \mu_0 \Omega^2 D(\mathbf{r},\mathbf{r}^\prime, \Omega)/(2\pi)  $ and $C(\mathbf{r},\mathbf{r}^\prime, \Omega) = \hbar \mu_0 \mathrm{sgn}[\Omega]\Omega^2 \mathrm{Im}[D(\mathbf{r},\mathbf{r}^\prime, \Omega)]/(2\pi)$, respectively, with the $xx$ component of the dyadic Green tensor of the vector Helmholtz equation $D$ and the vacuum permeability $\mu_0$.  

We further follow Ref.~\cite{Lindel2023ProbingTime} to numerically evaluate Eqs.~\eqref{eq:EOSSignalVF} and \eqref{eq:EOSSignalSR}. The resulting frequency domain correlation signals defined by $G_i(\Omega) = (1/2\pi)\int_{-\infty}^\infty \mathrm{d}\delta t G_i(\delta t)\me^{\mi \Omega \delta t}$ are shown in Fig.~\ref{fig:results}c. We used the following parameters in the simulation, which, except for $n_g$, have all been independently determined experimentally:
pulse duration $\tau_\mathrm{FWHM} = 110\,$fs, Gaussian beam width $w = \SI{10}{\um}$, beam separation $\delta r =\SI{50}{\um}$, and the group refractive index in the near-infrared $n_g = 3.18$. The refractive index in the THz is given by
\begin{align}
    n(\Omega) = \sqrt{ \epsilon_\infty  \left(1 + \frac{(\hbar \omega_\mathrm{LO})^2 - (\hbar \omega_\mathrm{TO})^2}{(\hbar\omega_\mathrm{TO})^2 - (\hbar\Omega /2\pi)^2-\mathrm{i}\hbar \Omega \gamma/2\pi  } \right)  },
\end{align}
with all parameters as measured in Ref.~\cite{Leitenstorfer1999DetectorsTheory} for a ZnTe crystal at temperature $T = 300\,$K, except that we use $\epsilon_\infty = 7.38$ to match the real part of the refractive index measured at $T = 4\,$K reported in Ref.~\cite{Settembrini2022DetectionCone}.

The fluctuation dissipation theorem implies $C(\mathbf{r},\mathbf{r}^\prime, \Omega) = \hbar\,\operatorname{sgn}(\Omega) \mathrm{Im}[R(\mathbf{r},\mathbf{r}^\prime,\Omega)]$, leading to the relation between $G_\mathrm{s}(\Omega)$ and $G_\mathrm{vac}(\Omega)$ discussed in the main text \cite{Lindel2023ProbingTime}.

\subsection*{Data availability statement}
The acquired correlation data that support the findings of this study are available in the ETHZ Research Collection with the following DOI: 10.3929/ethz-b-000723846

\section*{End notes}

\subsection*{Acknowledgements}
 A.H. acknowledges financial support from the Swiss national foundation, project 200021-21273. F.L. acknowledges support by the Spanish Ministry for Science and Innovation—Agencia Estatal de Investigación (AEI) through Grant No. EUR2023-143478, as well as by the Quantum Center Research Fellowship and the Dr. Alfred and Flora Spälti Fonds.

\subsection*{Author contributions}
J.F., A.H. and F.L. conceived the idea for the experiment and its theoretical interpretation. A.H. and L.G. modified the experimental setup. A.H. conducted the measurements. The data analysis was primarily performed by A.H. and their results were interpreted by A.H., F.L. and J.F. The theoretical framework was developed by F.L. and S.Y.B. F.L. performed the numerical simulations. J.F. was the scientific supervisor of this work. The manuscript was written through contributions from all authors. All authors have given approval to the final version of the manuscript. 

\subsection*{Competing interest declaration}
The authors declare no competing interests. 

\subsection*{Additional information}
Supplementary Information is available for this paper. Correspondence and requests for materials should be addressed to Alexa Herter (aherter@ethz.ch), Frieder Lindel (flindel@ethz.ch) or Jérôme Faist (jfaist@ethz.ch). Reprints and permissions information is available at www.nature.com/reprints. 
\subsubsection*{List of Extended data figures}
\begin{enumerate}

    \item  \textbf{Pulse length inside detection crystal}\\
    Measured intensity auto-correlation of probe signal inside ZnTe-crystal: The detected date points are interpolated with a cubic spline. The full-width half-maximum of the intensity auto-correlation for the optimized prism position measures \SI{168}{\fs} corresponding to a pulse length of \SI{110}{\fs} for a sech$^2$-shaped pulse. 
    \item \textbf{Extraction of zero-time delay in experimental data}\\
    \textbf{a:} Stage position, where the non-RF referenced correlation acquired in the configuration sensitive to vacuum fluctuations plotted for the individually scanned temporal correlation traces. This peak indicates the stage position, for which the two probe pulses have zero time delay. \textbf{b:} Fitting the complex phase of the vacuum-induced spectral correlation to extract the remaining shift of the zero-time delay. The gray area indicates the region not considered for the fitting due to the low amplitude of the correlation. \textbf{c:} Real and imaginary part of vacuum-induced spectral correlation for corrected spectral phase. The shaded area indicate the phase-uncertainty corresponding to a temporal shift of $\pm \SI{16.7}{\fs}$ or the time distance between two neighboring data-points. \textbf{d}: Temporal correlation induced by source radiation and filtered with a \SI{5}{\THz} low-pass to suppress high-frequency noise. Two different measuring sets have been averaged and a temporal shift between the two sets is observable. \textbf{e:} Cross-correlation between the two temporal correlations shown in panel \textbf{d}. \textbf{f:} Real and imaginary part of spectral correlation induced by source radiation after correcting the complex phase. Similar to panel \textbf{c}, the phase-uncertainty is indicated by shaded areas.
    \item \textbf{Influence of beam distance on vacuum-induced correlation}\\
    \textbf{a:} Measured spectral correlation induced by vacuum fluctuation compared to simulated results assuming different beam distances between \SI{50}{\um} and \SI{30}{\um}. \textbf{b:} Vacuum-induced temporal correlation compared to simulation assuming a beam distance of \SI{30}{\um}.
\end{enumerate}

\end{document}